# A thematic analysis of highly retweeted early COVID-19 tweets: Consensus, information, dissent, and lockdown life[1]

Mike Thelwall, Saheeda Thelwall, University of Wolverhampton.

**Purpose**: Public attitudes towards COVID-19 and social distancing are critical in reducing its spread. It is therefore important to understand public reactions and information dissemination in all major forms, including on social media. This article investigates important issues reflected on Twitter in the early stages of the public reaction to COVID-19.
**Design/methodology/approach**: A thematic analysis of the most retweeted English-language tweets mentioning COVID-19 during March 10-29, 2020.
**Findings:** The main themes identified for the 87 qualifying tweets accounting for 14 million retweets were: lockdown life; attitude towards social restrictions; politics; safety messages; people with COVID-19; support for key workers; work; and COVID-19 facts/news.
**Research limitations/implications:** Twitter played many positive roles, mainly through unofficial tweets. Users shared social distancing information, helped build support for social distancing, criticised government responses, expressed support for key workers, and helped each other cope with social isolation. A few popular tweets not supporting social distancing show that government messages sometimes failed.
**Practical implications**: Public health campaigns in future may consider encouraging grass roots social web activity to support campaign goals. At a methodological level, analysing retweet counts emphasised politics and ignored practical implementation issues.
**Originality/value**: This is the first qualitative analysis of general COVID-19-related retweeting.

# Introduction

In the COVID-19 crisis, the social sciences have an important part to play in this disease because a core part of public health strategies defence is social distancing (e.g., Tian, Liu, Li, et al., 2020). This relies on communicating the necessary behaviour effectively to the population and engaging their consent to comply. Effective public communication is therefore critical. Moreover, public reactions to events need to be assessed to understand the mediums in which information and consensus spread effectively. The public perspective is therefore more central for this than for most other major public health threats in the past century. Information about public behaviours or attitudes may also help understand the epidemiology of the disease, which is important to help control it (Lipsitch, Swerdlow, & Finelli, 2020).

Social media can help to inform and connect the population, bringing social support during lockdown (e.g., Wiederhold, 2020). Twitter is an important source of breaking news for a minority of people in some Western countries (e.g., the USA: Shearer, 2018). It can also be an important vehicle for disseminating new public health information (Liang, Fung, Tse, et al., 2019), although it usually relies on retweeting to attract a large audience (Steele & Dumbrell, 2012). During a crisis, members of the public can be expected to tweet about their own situation and emotional reactions (Lachlan, Spence, Lin, Najarian, & Del Greco, 2016). These tweets can be exploited to help public health officials to understand the situation on

---

the ground. For example, tweets related to the Zika virus have been analysed to identify the main public concerns (Glowacki, Lazard, Wilcox, Mackert, & Bernhardt, 2016), and which messages were retweeted to the largest audience (Stefanidis, Vraga, Lamprianidis, et al., 2017). For Ebola, Twitter seemed to serve as a filter for topics with media coverage (Morin, Bost, Mercier, Dozon, & Atlani-Duault, 2018). Perhaps surprisingly, preventative measure information is not always the most widely retweeted (Vijaykumar, Nowak, Himelboim, & Jin, 2018). For COVID-19, some insights have already been gained from Twitter: females are more likely to discuss social distancing and males are more likely to tweet about sport cancellations (Thelwall & Thelwall, 2020), misinformation and valid information spread in similar ways (Cinelli, Quattrociocchi, Galeazzi et al., 2020) and tweeters with wider networks are less impolite when discussing the disease (Kim, 2020).

Retweeting is an important aspect of the social media information ecology (Alhabash & McAlister, 2015). Retweeting has previously been used to gain insights into public attitudes (e.g., Dare-Edwards, 2014; Gabarron, Makhlysheva, & Marco, 2015; McNeil, Brna, & Gordon, 2012) or social movements, as evidence of the effectiveness of official information dissemination strategies (e.g., for Hurricane Sandy: Wang & Zhuang, 2017), or as evidence of the importance of the retweeted content (e.g., Hermida, 2013; Starbird & Palen, 2012; David, Ong, & Legara, 2016). During the Japanese earthquake of 2012, tweets from people in the affected area were retweeted by those outside to share first-hand information quickly (Miyabe, Miura, & Aramaki, 2012). This illustrates that tweets from citizens (rather than politicians, experts, or celebrities) can reach large audiences during crises, especially when they have first-hand accounts or information to share.

Studies of political retweeting give some insights into factors influencing success. Official sources are not always the most successful (Sanjari & Khazraee, 2014; Segesten, & Bossetta, 2017), there may be geopolitical differences in the types of tweets that are most retweeted, humour seems to drive much retweeting (Driscoll, Ananny, Bar, et al., 2013) and bots can be problematic (Stewart, Arif, & Starbird, 2018). In normal times, a tweet is more likely to be retweeted if it contains a URL or hashtag, and if the tweeter has many followers and follows many people (Suh, Hong, Pirolli, & Chi, 2010), although this may have changed after 2017, when the character limit was doubled from 140 to 280.

For Covid-19, retweeting is more common for scientific information than for false facts (Pulido, Villarejo-Carballido, Redondo-Sama, & Gómez, 2020). For disability issues, highly retweeted tweets have been found to offer support, explain problems and reflect outrage at people dismissing the value of people with disabilities (Thelwall & Levitt, 2020). Twitter is also used to share academic research about Covid-19, with scholarly tweeters' contributions frequently attracting retweets (Fang & Costas, 2020). The overwhelming majority (89%) of eight G7 world leaders' viral tweets about COVID-19 by March 17, 2020 were primarily informative, emphasising the potential role of politicians in disseminating information (Rufai & Bunce, 2020).

Not everyone engages with social media sites, and Twitter is used by a minority of the population of any country. In the USA, 22% of adults use Twitter and the most active tweeters are more likely to be female and to post about politics (Wojcik & Hughes, 2019). More educated Twitter users are more confident to share information and to use the site for pleasure (Syn, & Oh, 2015). Nevertheless, during a crisis people that normally tweet everyday events may switch to tweeting crisis-related information (Palen, Starbird, Vieweg, & Hughes, 2010). Information sharing and consumption on the social web is influenced by participants' lifeworld background as well as the norms of the small world formed by the people that they interact with (Burnett & Jaeger, 2008). In a political context, likeminded groups can form that discuss issues within their collective worldview, away from mainstream opinions (Fraser, 1990). Because of these factors affecting who tweets and what topics they tweet about, it is not possible to make strong claims about society from tweets. Instead an investigation of Twitter can show what is happening within its ecosphere and make suggestions about how it may have been influenced by wider societal issues. This is a major limitation for the current paper, but Twitter is nevertheless worth investigating for Covid-19 because it provides a uniquely public large-scale source of public reactions and so the suggestions from any Twitter study may provide unique perspectives. It may also provide evidence that can be triangulated with other studies to increase consensus about core public interest issues around Covid-19.

This article investigates highly retweeted tweets during March 2020 about COVID-19 for insights into public reactions to the pandemic in its early stages and the role of Twitter in information spreading and consensus building. A highly retweeted tweet for COVID-19 will have been seen by many people and may therefore be influential. It is also is likely to have resonated in some way to motivate sharing, and therefore may shed light on an aspect of public attitudes to COVID-19, at least as mediated by the Twitter ecosystem. As the above brief review suggests, there is no expected type of tweet that may become highly retweeted, but the set is likely to include citizen tweets.

## Public health campaigns

The informational component of the COVID-19 reaction has centred on public information campaigns in a health context. For a public health campaign to be successful, the message must be seen and acted upon. Between these two stages, recipients usually need to understand the message, believe that it applies to them and will benefit them or their community, have the will to carry out the necessary actions, and then translate this will into action. A break in any of these stages is likely to lead to an initiative failing (e.g., Passmore, Williams-Parry, Casper, & Thomas, 2017). The situation for COVID-19 has been different in that behaviour change is mandated by government, but it still relies on the public understanding the message and carrying it out effectively.

### *Global epidemics*

Previous studies of health promotion strategies for global epidemics have few insights that would help an analysis of public reactions to COVID-19, and so are only discussed briefly here. For example, the Spanish flu pandemic of 1918-19 was addressed with social distancing measures (Caley, Philp, & McCracken, 2008; McCracken & Curson 2003) but communication measures used then may have little relevance today.

The SARS (severe acute respiratory syndrome) coronavirus epidemic of 2003 was contained primarily through finding infected patients, tracing their contacts and isolating them to reduce the transmission rate, although social distancing and mask wearing were also used (Bell, 2004; James, Shindo, Cutter, Ma, & Chew, 2006). The social distancing measures included school closures and cancelling mass events. The relatively rapid containment of the disease meant that it was not politicised in the news (Lewison, 2008). The role of social media has been examined for SARS from the perspective of health organisations disseminating their messages (Guidry, Jin, Orr, Messner, & Meganck, 2017), but this is not directly relevant here because it does not give insights into issues that resonated with the public. For Ebola, an analysis of retweeting found that tweets typically had little depth for retweeting, with 91% of retweets being from the original tweet (Liang, Fung, Tse, et al., 2019). Thus, prior research has little relevance for the current study.

*Smoking bans*

Knowledge from previous health promotion interventions and research can inform responses to COVID-19 (Van den Broucke, 2020). In particular, national strategies to reduce second-hand exposure to tobacco smoke provide a good starting point than global epidemic research for the current study. Like COVID-19 Reponses, they involve legal enforcement but need widespread partly voluntary compliance to be effective. These strategies have been extensively and systematically researched, providing detailed insights into the factors associated with success or compliance. In particular, a recent international qualitative systematic review found seven themes in studies of the barriers to effective implementation of smoke-free homes (Passey, Longman, Robinson, Wiggers, & Jones, 2016), which form a useful benchmark for the current study. Smoking-related issues most relevant to pandemics are highlighted below.

- *Knowledge, awareness and risk perception*: This includes risk denials, differing perceptions of acceptable risk levels, and differing engagement with the need to protect others' health.
- *Agency and personal skills/attributes*: The environmental constraint of the need to share living space affected household policies.
- *Wider community norms and personal moral responsibilities*: This includes the extent to which community responsibility is a cultural norm and the extent to which responsible actions are the community norm.
- *Social relationships and influence of others*: Individuals within a family may trigger family-wide discussions about acceptable practices.
- *Perceived benefits, preferences and priorities*: Positive side-effects of enforcing anti-smoking rules in a household helped to start and maintain them.
- *Addiction and habit*: Habituation was a barrier to behaviour change.
- *Practicalities*: Inconvenience was a barrier to smoke-free homes unless practical solutions could be found.

These seven themes all seem relevant to COVID-19, except perhaps for addiction and habit. To illustrate some of the factors: clear information about the harms of second-hand smoke is important for successfully setting up smoke-free homes (Milcarz, Bak-Romaniszyn, & Kaleta, 2017; Stevenson, Campbell, Gould, Robertson, & Clough, 2017); families negotiate around micro-level concerns to agree constraints to limit their children's smoke exposure (Poland, Gastaldo, Pancham, & Ferrence, 2009); and the greater prevalence of smoking amongst lower socio-economic groups in the USA is partly due to fewer being able to live in smoke-free homes (Vijayaraghavan, Benmarnhia, Pierce, et al., 2018).

## Interpretation of retweet counts

Social media can help reach a wide audience (Korda & Itani, 2013) and retweeting can be used as an indicator of moderate engagement with a public health campaign (Neiger, Thackeray, Burton, Giraud-Carrier, & Fagen, 2013). The current article analyses tweets that have been retweeted at least 100,000 times, ignoring the presumably low percentage of retweets by bots (since this is not a primarily political topic). There are many reasons why someone might retweet, depending on the nature of the tweet and reliability of the original tweeter and the interests of the tweeter's followers (Engelmann, Kloss, Neuberger, & Brockmann, 2019; Metaxas, Mustafaraj, Wong, Zeng, O'Keefe, & Finn, 2015; Rudat, Buder, & Hesse, 2014). These reasons include at least the following topic-related factors (Boyd, Golder, & Lotan, 2010; Lee, Kim, & Kim, 2015; Metaxas et al., 2015).

- A belief that the information will be interesting, useful or entertaining.
- To persuade others.
- To associate with the message, such as by showing agreement.
- To reward the tweeter for positive content.
- To be a visible part of a conversation.
- To save the tweet for reference.
- In the hope of reciprocal retweets.

The mix of reasons for retweeting varies based on the issue concerned and may be counter-intuitive. For example, an analysis of Breast Cancer Awareness Month tweets found that the most retweeted were promoting the Month or fundraising rather than educational (Chung, 2017). In contrast, highly retweeted tweets about Hurricane Irma were likely to contain pictures, safety instructions or information (Lachlan, Xu, Hutter, Adam, & Spence, 2019).

There are multiple reasons for using Twitter, including information, socialising and entertainment (Alhabash & McAlister, 2015; Liu, Cheung, & Lee, 2010). There are also multiple corresponding causes of high retweet counts. Nevertheless, each highly retweeted tweet potentially gives an insight into an issue that resonates with the public. For a useful insight, the reason for the tweet must be inferred from its texts and context, which entails a qualitative judgement. Some reasons may be irrelevant or obvious, but others might give new insights. Thus, a multi-stage approach is needed to extract useful meaning from retweets.

1. **Collection**: Collect a relevant collection of tweets and identify a subset that is highly retweeted.
2. **Typology**: Make a subjective judgement about the type of information contained or the reason for retweeting in each case.
3. **Insights**: Match the types/reasons identified for highly retweeted tweets with what is known about the issue and identify those that represent new insights into the issue.

In comparison to typical qualitative research, qualitative retweet analysis is quick for data collection and analysis (because of the lower number of texts). Moreover, there is large sample support for each individual retweet, reducing the risk that findings apply to only a small group, as in typical qualitative case studies.

For COVID-19, there are specific sources of bias on Twitter. In the UK, a government team is targeting fake news on social media, such as modifications of its core messages, "GOV.UK CORONAVIRUS ALERT New rules are in force: you must stay at home. More info & exceptions at gov.uk/coronavirus Stay at home. Protect the NHS. Save lives." (BBC, 2020). This message was texted across the UK and tweeted by the UK prime minister as a picture message on March 24 but had been retweeted only three thousand times by 30 March, 2020[2]. From March 16, Twitter had also been removing misleading advice that might have fatal consequences, in consultation with governments and other relevant organisations (Twitter, 2020). Thus, any set of highly retweeted tweets is censored, excluding widely believed but misleading posts. For example, scams, alternative medicine and religious messages advocating unsafe practices or ineffective cures could expect to be taken down.

---

[2] https://twitter.com/10DowningStreet/status/1242412113127190531

# Methods

**Collection**: A topical collection of tweets was collected from the Twitter Search API (Applications Programming Interface) using the free software Mozdeh between 10 and 29 March 2020 with the following queries: coronavirus; COVID-19; COVID19; "corona virus". The first three queries match the term irrespective of whether it is a hashtag. Testing with Twitter online and pilot testing with the API suggested that these would capture a substantial number of relevant tweets. Other query terms, such as #socialdistancing or #WuhanVirus, could have captured additional relevant tweets but a set that is specific to the virus rather than an aspect of the reaction to it gives a general collection that is not biased towards any particular reaction. Nevertheless, the set of queries used contains its own biases, such as towards people that use relatively precise terms, in contrast to, for example, "the corona" or "the virus". It also omits many discussions of COVID-19 that do not explicitly mention it (e.g., "can't see my boyfriend until this ends").

The queries produced 23,603,317 matching tweets. This set was processed to identify all tweets with a retweet count of at least 100,000, as reported by the API. These were separated for analysis and ranked in descending order. Duplicate tweets (either identical text or one being a truncated version of the other) were deleted, as were tweets originally posted before March 10, 2020. The duplicate elimination here focuses on the text of the tweets. If a more technical criterion had been used, such as tweets with the same ID, then the results could have included multiple tweets with the same text, which would not have given additional information. Mozdeh did not attempt to merge the retweet counts for multiple copies of identical tweets, so it is possible that some popular tweets were not included in the dataset because separate copies were not retweeted enough to be included even though their combined retweet count would have been high enough.

**Typology**: Each of the top tweets was located on Twitter.com and read within the site to identify the full tweet (the API truncates long retweets) and check any embedded links or images. Using a thematic analysis approach (Braun & Clarke, 2006), the tweets were tagged with one or more descriptions, then repeatedly clustered and re-classified to generate a set of consistently applied tags reflecting the main purposes of each tweet. This process was conducted independently by both authors, then combining the themes and subthemes to produce a shared final set. This is a qualitative and subjective approach. This coding stage was conducted before the literature review and was not guided by the smoke free homes research, which was unknown to the authors at the time of coding. The second author is an experienced health promotion facilitator, nurse and nurse educator with a wide awareness of health promotion issues. This will have influenced the coding to some extent.

Many tweets involved humour. Although there are many current psychological theories of humour, two prominent ones emphasise the social context of a joke as central to its humour. Reversal theory (Apter, 1989) argues that a joke's context must increase physiological arousal in either positive or negative ways and then diminish it. The seriousness of COVID-19 might be one such context. For example, a joke centring on the threat of the virus would not be funny and therefore not retweeted by people that were not alarmed by COVID-19, so high retweets would be evidence of public perception of COVID-19 as threatening. Similarly, benign violation theory (McGraw & Warren, 2010) argues that humour arises when something is simultaneously threatening and non-threatening. For example, a threat expressed by paraphrasing a song lyric could be funny due to the song lyric playing the non-threatening role. A joke related to COVID-19 being funny therefore implies that an aspect of the joke was perceived to be threatening. Thus, when relevant to COVID-19, jokes were classified according to the underlying message rather than primarily as a joke.

**Insights**: Insights were identified in three stages. First, each theme and, when necessary, each tweet was evaluated for information relevant to COVID-19 that could be inferred from the high retweet status. This entailed judging the likely common motivation for retweeting. Second, motivations not directly relevant to COVID-19 as a public health issue were discarded (e.g., humour in one case). Third, the remaining motivations were assessed in the context of prior research and assessed for (a) confirming expected patterns or (b) new insights. The insights are reported in the Discussion section. Some of the retweets may have been created by bots but a study of COVID-19 has found bots to be most prevalent for conspiracy theories (Ferrara, 2020), which were not present in the data set. Astroturfing, the inflation of retweet counts through co-ordinated or automatic retweeting, seems to be most common for political tweeting (Keller, Schoch, Stier, & Yang, 2020), so it is possible that the political tweets had inflated counts, but this is difficult to detect because astroturfing imitates human tweeting and may be conducted by paid or volunteer humans. All the political tweets seemed to be plausibly popular, however. To test for this, at least twenty quote tweets and bios of tweeters that had liked or retweeted were examined for each political tweet, as reported by Twitter on the tweet homepage. This manual check did not reveal evidence of systematic astroturfing in the quote tweets, which contained comments on the original tweets that were varied and plausible, even for the most political tweets (e.g., directly anti-Trump). Some retweeters and likers could have been fake, and some were accounts for sex workers or advertising, perhaps retweeting to attract an audience. It is therefore likely that some of the retweet counts were inflated. Nevertheless, this seems to happen for many (perhaps all) popular tweets and the large number of plausible quote tweets gives evidence that all popular political tweets genuinely resonated on Twitter.

**Ethics**: Tweets are not private but are fully in the public domain and therefore do not need ethical approval to research at the University of Wolverhampton, although informed consent is needed before including quotes because these might generate unwanted publicity (Eysenbach & Till, 2001; Golder, Ahmed, Norman, & Booth, 2017; Wilkinson & Thelwall, 2011). In the current study, although all tweets have been seen by at least 100,000 tweeters, some were personal stories about the death of a family member, and many were tweeted by people without many followers. Thus, no exact quotes were included for the tweets and no list of tweets or Twitter IDs is included. Heavily modified pseudo-quotes are included in brackets (rather than quotes) and subheadings to give a flavour of the results, instead.

# Results

After merging duplicates, 87 different English-language tweets posted on or after March 10, 2010 and containing coronavirus, COVID-19, COVID19, or "corona virus" had received at least 100,000 retweets by March 29, 2020. Collectively, they accounted for 14 million retweets and probably about twice as many Likes. The themes and subthemes identified are reported below in descending order of the total number of retweets.

## Lockdown life: Penguins loose in the Chicago aquarium

Twenty-eight tweets (4.2 million retweets) discussed various aspects of life under self-isolation, lockdown or social distancing rules. Six (1.2m) discussed the situation of **toddlers or animals** during the lockdown, using cute videos or jokes (a penguin release story was apparently real: Guardian, 2020). Four (0.6m) discussed **community-focused lockdown**

**activities,** such as community singing or helping people that could not afford to pay rent, one of which was humorous, and one (0.1m) tweeted a humorous picture of innovative solo timewasting. Three (0.3m) discussed what people might do **after the lockdown** ends (will we be as hygienic?). Two (0.3m) discussed **event cancellations** with crowds, both with humour. For example, one tweeted that a fictional race had been cancelled, the hyperbole presumably working on the realisation that everything had been cancelled. Three tweets (0.3m) were jokes about the **virus dominating all news and conversation** (when will we talk about me instead?) and three (0.5m) joked that **extreme social distancing measures might be necessary** (e.g., dogs fetching shopping).

The remaining tweets discussed different topics: wanting the lockdown to **last at least two months** (0.2m), fear of panic buying (033m), joking that some people were getting their core **information from friends** (0.1m), **missing friends** (0.1m, cute video), **fear of testing** joke (0.1m), and a **tasteless joke** about old people dying (0.1m, from March 12).

## *Attitude: COVID-19 jokes are funny but I'm still frightened*

Eighteen tweets (2.6 million retweets) primarily expressed an attitude towards the COVID-19 pandemic restrictions. Eight tweets (1.2m) emphasised the **threat or seriousness** of the pandemic, all in the form of jokes or humour (e.g., a graphic of a city with a wall around it proposed as a solution). In contrast, three tweets (0.4m), one in the form of a joke, inferred **dissent** about the need for social distancing (e.g., I won't let it stop me going on holiday because COVID-19 is in every country). Between these two, five tweets expressed **hope** about personal protection (2 tweets, 0.4m), both jokes (e.g., the drugs already in my system will beat it), or for an end of pandemic (3 tweets, 0.3m), one joke (K-pop fans will cure it if it infects their idol), and one religious (e.g., I believe Jesus will end it soon). Finally, two tweets (0.3m) expressed **dismay** at the lockdown restrictions (e.g., all my plans are cancelled).

Partly conflicting opinions were expressed on Twitter but the apparently dominant attitude of fear was primarily expressed through humour, whereas the opposite was mainly expressed seriously. More sceptical tweets tended to be early (March 10-14), although the same was true for the threat tweets (exception: March 18). One of the dissent tweets had been removed at the time of testing, perhaps by the owner receiving negative feedback or as a Twitter.com safety action.

## *Political: Katie Porter wins free coronavirus testing*

Fifteen tweets (2.2 million retweets) discussed political issues, primarily focusing on the USA, and none containing humour. Five tweets (0.9m) praised or criticised **politicians' actions**, two accusing President Trump of mismanagement, two praising or announcing Representative Katie Porter getting a free coronavirus testing promise from the US administration, and one attacking Senator Richard Burr for putting profits first. Five tweets (0.7m) **attacked (mainly US) government policy** (before the outbreak) about healthcare, human rights, and spending, based on COVID-19-related events. Three tweets (0.4m) **accused business of greed or dishonesty** before the virus (not allowing workers to stay at home) or during it (putting profits before lives). Two other political tweets praised **Cuba** for helping Italy or accused people of **racism** for calling COVID-19 a Chinese virus.

## *Safety messages: Stupid will kill us*

Eleven tweets (1.9 million retweets) gave safety information, such as the need for social distancing or obeying government lockdown rules. Eight (1.3m) were straightforward and relatively serious **exhortations to follow social distancing guidelines**. These tweets did not include or explain the guidelines but attempted to persuade that they were important (if you childishly don't follow instructions, we'll all suffer). An additional three (0.6m) embedded specific **social distancing advice** in humour in the form of a comic video message (featuring Mel Brooks) and two song lyrics rewritten for social distancing advice (Queen's Bohemian Rhapsody; Dua Lipa's Don't Start Now). These did not give comprehensive social distancing instructions but implicitly or explicitly highlighted the need for it, as well as mentioning its nature.

### *People with COVID-19: It was like this for me*

Eight tweets (1.5 million retweets) discussed individuals who had caught the disease. Three reported **celebrities testing positive** (0.6m), two as personal announcements (Boris Johnson, Idris Elba), and one in the form of a joke. Three further tweets reported in detail (Twitter threads) the **experience of COVID-19 symptoms** for the tweeter, who had just recovered (0.6m). Two (0.3m) reported the **impact of a COVID-19 death** (of a parent) on the tweeter, combining it with an exhortation to follow social distancing rules.

### *Support key workers: Nurses, doctors, cleaners, shopworkers are the vital people*

Four tweets (0.5 million retweets) were **messages of support for all key workers** during the crisis (two tweets, 0.3m) or **messages of support for medics** (2 tweets, 0.2m). These seemed to be partly expressing appreciation and partly making a political point (not bankers or traders, it's the nurses and cleaners that are saving us).

### *Work: Meetings by email*

Three of the most retweeted tweets (0.5 million retweets) discussed separate work-related issues: the discovery that meetings can take place by email (0.2m), an ineffective workplace preventative measure against COVID-19 (0.1m, possibly a joke) and a criticism of social media influencers that were shaming people for working despite the restrictions, since they might need the money to live (0.1m).

### *COVID-19: Daily symptoms*

Only two of the most retweeted tweets (0.3 million retweets) focused on the virus. One described the **symptoms** day by day with an animation showing the impact of the virus on the body (0.3m) and the other was a **news** story about a fast COVID-19 test (a genuine news story but false).

## Discussion

In addition to the limitations discussed above, this study focuses on information rather than behaviour. It is possible, for example, that people widely retweet public safety information and then ignore it. This seems unlikely, however, and one previous study of Zika virus concerns has shown a connection between public attitudes and the volume of tweets about Zika (Farhadloo, Winneg, Chan, Jamieson, & Albarracin, 2018), supporting the claim that Twitter can influence public health behaviour. The study is also limited in its focus on highly retweeted tweets. Other issues may well have been more extensively discussed on Twitter than some reported below but lacking individual highly retweeted tweets. The analysis also makes inferences about the purpose or impact of the tweets, which may not be correct. In particular, memetic qualities of jokes may have been missed, with the resulting associations changing their meaning (Shifman, Levy, & Thelwall, 2014). The results should be interpreted as applying to the demographic of Twitter users, which is a minority of the USA and presumably excludes the very young and the very old, for example, and people interested in politics may be overrepresented. Thus, and in conjunction with possible astroturfing, the importance of political issues may be exaggerated in the findings.

The tweets are from the early stage of the public reaction to COVID-19 and later discussions may well focus on different issues. For reference in terms of the period examined (March 10-29), on March 18 the UK government advised people with symptoms to self-isolate and advised everyone to avoid unnecessary gatherings and to work from home if possible. It implemented a nationwide stay-at-home order on 23 March. In the USA, stay-at-home orders were implemented separately by each state or region, starting in some parts of California on 17 March, becoming state-wide two days later. A New York state lockdown started on 20 March and other states joined afterwards, including eight on the 23rd of March. Many US states had not implemented a stay-at-home order by March 29, however.

In terms of cross-cutting themes, the prominence of joke tweets reflects the common use of Twitter for entertainment (Alhabash & McAlister, 2015; Liu et al., 2010). Humour almost always accompanied a COVID-19 message, although it is difficult to judge whether the message played a subordinate role in the retweeting. Humour seems likely to boost the audience of a tweet by making it more likely to be shared, however. Other cross-cutting themes include human interest stories and political points (including outside the politics theme). Politics is a known international use for Twitter (Ozturkcan, Kasap, Cevik, & Zaman, 2017). This also highlights the fact that tweets may become highly retweeted for multiple reasons, including some that are not obvious from a surface reading. These factors include the number of followers of the original tweeter, the match of the tweet to the message, external factors (e.g., if the tweet was promoted in the news), the pithiness of the text, timeliness, and any human interest angles. Nevertheless, it seems unlikely that a tweet would become highly retweeted without having a message that did not resonate with many people, unless to share disgust at the tweeter (which did not seem to be the case in the sample).

## *Lockdown life*

The 4.2 million retweets about quarantine, lockdown or social distancing rules reflect a rapid and substantial change in daily lives. The jokes and humorous videos may help people to come to terms with changes in their lives. Presumably the changes resulted in many sources of uncertainty and stress, which humour could partially alleviate, at least in the short term (Martin & Ford, 2018). In Italy, a study of Covid-19 humour found that people less at risk were more likely to find jokes funny (Bischetti, Canal, & Bambini, 2020), so there may be space for humour on Twitter amongst those feeling least threatened. The absence of tweets about real sports cancellations, in contrast to an earlier study of typical tweets (Thelwall & Thelwall, 2020), suggests that mass cancellations were more important than individual events.

## *Attitude*

It is unsurprising that 2.6 million retweets expressed an attitude towards the COVID-19 restrictions, given their substantial negative impact on many lives. The tweets reveal substantial support for different opinions. Although accepting the issue as a serious threat was the dominant attitude (also evident in the safety message retweeting), this was only conveyed with humour. It is possible that those taking the outbreak most seriously tended to retweet a more proactive tweet, such as a safety message

Popular tweets expressing dissent confirm that the population was divided about the issue and that **initial government safety messages were not universally persuasive**. As a practical implication, tracking Twitter might help governments to assess the level of credibility of their public health messages, as they presumably already do, in order to target advertising or campaign messages.

### *Political*

The 2.2 million retweets focusing on political issues were all anti-establishment in the sense of being critical of the current national government. This might reflect the demographics of Twitter users rather than broader society, however. It seems likely that it is more used as an alternative news or politics channel by people that disagree with the mainstream media, for example. These tweets might play the important democratic function of holding the government to account and creating pressure on the government to carry out specific actions, or keep to its promises (e.g., on free testing in the USA). Political discussions on Twitter seem to mainly occur between likeminded people rather than being deliberations towards a consensus (Lorentzen, 2016). Thus, the highly retweeted tweets should not be interpreted as consensus opinions.

### *Safety messages*

The 1.9 million retweets for non-official messages offering direct exhortation or indirect encouragement to follow safety rules represent a **substantial free resource to spread information about social restrictions** and, perhaps more importantly, to **create community support** for following them. The latter may be more important in the US and UK, for example, with the safety advice having been given by politicians (Trump, Johnson) that are distrusted by at least a substantial minority of the population. This support presumably helped to build compliance with the otherwise draconian measures. Support message seem likely to have been most effective when the message was purely serious since there would be little possibility to misinterpret it as entertainment.

The humour-related messages also reached a wide audience and their role may have been more informational for the reasons above. Celebrity endorsements (e.g., Mel Brooks) may well allow them to contain a persuasive element, through parasocial connections, respect or shared community membership. Humorous messages may also reach a different audience, those using Twitter primarily for entertainment rather than other purposes, such as news.

As a caveat, retweeting a safety message does not necessarily mean that the retweeter followed the advice or that their friends were persuaded by it. Moreover, even if someone felt persuaded then this would not necessarily transfer into behaviour changes (Cugelman, Thelwall, & Dawes, 2011). Thus, overall, safety messages were widely shared informally on Twitter, reaching a large audience quickly with a personal connection, and retweeting serious and celebrity-endorsed humorous messages seems likely to have **helped build community-wide support for compliance**.

### *People with COVID-19*

The 1.5 million retweets about individuals who had caught the disease represent substantial human interest, perhaps partly due to concern for the celebrities testing positive and partly due to fear of the disease, leading to a desire for information and reassurance about the symptoms. Detailed reports of survivor experiences seem likely to have been reassuring to those who might become infected, and informational to help them self-diagnose. Together with the reports of individual family member deaths, they seem likely to have served as safety messages, illustrating that anyone can be infected. The detailed experience reports included careful disclaimers that other people's experiences, symptoms and prognoses might be different. Thus, the personal stories seem to have been useful additions to the information ecology for COVID-19 for **safety messages** and **public health information**.

### *Support key workers*

The 0.5 million retweeted messages of support for all key workers or health professionals echoed public displays of gratitude in many countries (e.g., clapping on a pre-defined day and time). Both presumably helped to **boost the morale of key workers** (often low paid) performing critical and dangerous jobs.

### *Work: Meetings by email*

The 0.5 million retweets for work-related issues reflect the conflict created by the tension between needing to work (as an employee or employer) and public safety. They seem to **highlight the lack of clear guidelines for who should work and how** at some stages of the pandemic. There were no highly retweeted tweets with work-advice or working from home stories, although some lockdown life stories may apply to people working from home.

### *COVID-19*

There seemed to be relatively **little interest in detailed technical information about COVID-19** or testing technology (0.3 million retweets). It is surprising that stories by individuals were more retweeted than detailed explanations of the full range of symptoms that could be experienced, which seems to be more useful information. This is a tentative conclusion, since the detailed technical information might have been shared in an earlier tweet or in multiple similar tweets, each with lower retweet counts. It seems more likely, however, that people needing detailed information would seek it in official government websites rather than social media.

### *Comparison with smoke-free environment research*

As argued above, smoke-free environment research forms an appropriate benchmark to compare the results with. The following compares the main themes from this area (Passey, Longman, Robinson, Wiggers, & Jones, 2016) with the main themes above.
- *Knowledge, awareness and risk perception*: Represented in safety messages, individuals with COVID-19, and COVID-19 tweets.
- *Agency and personal skills/attributes*: Partly represented in the attitude theme.
- *Wider community norms and personal moral responsibilities*: Represented in the support for key workers to some extent.
- *Social relationships and influence of others*: Partly represented by the attitude theme.
- *Perceived benefits, preferences and priorities*: Not directly represented.
- *Addiction and habit*: Not directly represented, but minor relevance to COVID-19.
- *Practicalities*: Meetings by email represent this to some extent.

The political theme does not fit well into the above set, emphasising its relative importance for COVID-19 compared to other health campaigns, despite most seeming to have a political dimension. Although anti-smoking campaigns had a political dimension, politics seems to be more important for COVID-19.

The first issue above seems to be the most comprehensively covered by highly retweeted tweets. The remaining issues are only indirectly covered. For example, there are many known issues about the practicalities of COVID-19 lockdowns, such as domestic abuse risks, methods to deal with shared parenting for families living apart, households with vulnerable people but one member potentially exposed to COVID-19 at work, and the need for carers to protect the vulnerable, especially for those with multiple carers. Thus, issues that are of widespread relevance at a general level but are individualised in nature may not be found by retweet analysis. The omission of issues like this emphasises that analysing retweet counts will give little insights into many important issues, a methodological limitation with the emphasis on highly retweeted articles.

## Conclusions

This analysis has reported the main themes in 87 tweets, collectively accounting for 14 million retweets and probably about 30 million Likes. For people following COVID-19 on Twitter, the themes found may not be surprising, but the thematic analysis here adds structure and creates a long-term record of high profile initial COVID-19 reactions in English on this important social media site. This is important since the pandemic reaction evolved into different stages and so early events may be quickly forgotten.

From an information behaviour perspective, the political results are consistent with Twitter supporting groups that disagree with the dominant narrative in society (Burnett & Jaeger, 2008; Fraser, 1990). Nevertheless, the remaining themes seem to be general rather than specific to a group and hence suggest that popular tweets can break out from relatively self-contained groups within Twitter. More generally, the results confirm that during a crisis people will often tweet (in this case retweet) crisis-related information (Palen et al., 2010). The presence of humour has been observed on Twitter for some previous crises (Bruns, Burgess, Crawford, & Shaw, 2012; Spence, Lachlan, Lin, & del Greco, 2015), but not others (Palen et al., 2010) and is an under-explored dimension of crisis information sharing.

The results suggest that Twitter had positive impacts on COVID-19 in terms of sharing information, encouraging support for social distancing guidelines and helping people to cope emotionally with the changes, despite it also being used to share fake stories (BBC, 2020; Pulido, Villarejo-Carballido, Redondo-Sama, & Gómez, 2020). The results also suggest that more fundamental political conclusions were being drawn from government reactions, or that government opponents found many reasons to believe that their criticisms were vindicated or highlighted by COVID-19. This may be exaggerated by the demographic of Twitter users.

From a methodological perspective, a comparison of the themes found with a typology of issues known to be important for generating smoke-free homes found many gaps. Retweet analysis therefore seems likely to generate insights to a limited extent, systematically overlooking some types of issue, and particularly those with localised or individualised relevance. Thus, the relatively fast nature of retweet analysis should not be exploited at the expense of other methods.

Some suggestions for pandemic public information campaigns and information providers can be made, based on the above. These are evidence-based suggestions, in the sense of being derived from data, but are not proven or evidence-based conclusions because the data includes no ties to behaviour and is mediated by the Twitter ecosystem (both user bias and Twitter culture biases). Thus, the suggestions are insights for consideration by experts and, if possible, for more systematic evaluation with different methods in future research.

- Twitter, and other social media, should be considered in government information strategies since they can reach millions (at least). Arguably the most effective messages were serious tweets from citizens (rather than politicians) in support of the restrictions. Celebrity humorous support messages could also be effective. Future campaigns might consider how it is possible to ensure that such messages might spread early, such as by encouraging frontline health workers to report their experiences (if this did not already happen). Unofficial messages seemed to be more retweeted than official messages, although more people may have seen official messages.
- The presence of some widely retweeted messages undermining social distancing restrictions might be used as an indicator that they are not fully effective. Moreover, perhaps (if ethical) Twitter might consider targeting retweeters of these with additional encouragement to comply or encouraging people to persuade friends that disagree. This is in addition to Twitter banning misleading information.
- Twitter does not seem to be effective for spreading factual information about viruses, so campaigns to do this should focus elsewhere, or develop different Twitter-based strategies.
- Given the likely substantial psychological impact of the lockdown restrictions, social media might be recommended by health professionals as a way for people to partly overcome social isolation and, through humour, cope psychologically with their new situation.